\documentclass[superscriptaddress,floatfix,showpacs,balancelastpage,aps,pra,tightenlines,reprint]{revtex4-1}

\usepackage{amsmath}
\usepackage{amssymb}
\usepackage{textcomp}
\usepackage{graphicx}

\newcommand{\bra}[1]{\ensuremath{\left\langle #1\right|}}
\newcommand{\ket}[1]{\ensuremath{\left|#1\right\rangle}}
\newcommand{\expup}[1]{\mathrm{e}^{#1}}
\newcommand{\nvket}[2]{\ensuremath{\left|{#1}\right\rangle_{\text{e}}\left|{#2}\right\rangle_{\mathrm{n}}}}

\newcommand{\kete}[1]{\ensuremath{\left|#1\right\rangle_\mathrm{e}}}
\newcommand{\ketn}[1]{\ensuremath{\left|#1\right\rangle_\mathrm{n}}}
\newcommand{\brae}[1]{\ensuremath{\left\langle #1\right|_\mathrm{e}}}
\newcommand{\bran}[1]{\ensuremath{\left\langle #1\right|_\mathrm{n}}}
\newcommand{\egyro}[0]{\gamma_\mathrm{e}}
\newcommand{\ngyro}[0]{\gamma_\mathrm{n}}
\newcommand{\herm}[0]{\text{h.c.}}
\newcommand{\N}[1]{$^{#1}$N}
\newcommand{\Apara}[0]{\ensuremath{A_\parallel}}
\newcommand{\Aperp}[0]{\ensuremath{A_\perp}}
\newcommand{\nup}[0]{\ensuremath{\ket{\uparrow}_\mathrm{n}}}
\newcommand{\ndn}[0]{\ensuremath{\ket{\downarrow}_\mathrm{n}}}

\begin{document}
\title{High fidelity gate operations within the coupled nuclear and electron spins of a nitrogen vacancy center in diamond}

\author{Mark S.\,Everitt}\email{mark.s.everitt@gmail.com}
\author{Simon Devitt}
\affiliation{National Institute of Informatics, 2-1-2 Hitotsubashi, Chiyoda-ku, Tokyo 101-8430, Japan}
\author{W.\,J.\,Munro}
\affiliation{NTT Basic Research Laboratories, NTT Corporation, 3-1 Morinosato-Wakamiya, Atsugi, Kanagawa 243-0198, Japan}
\affiliation{National Institute of Informatics, 2-1-2 Hitotsubashi, Chiyoda-ku, Tokyo 101-8430, Japan}
\author{Kae Nemoto}\email{nemoto@nii.ac.jp}
\affiliation{National Institute of Informatics, 2-1-2 Hitotsubashi, Chiyoda-ku, Tokyo 101-8430, Japan}

\date{\today}

\begin{abstract}
In this article we investigate the dynamics of a single negatively charged nitrogen-vacancy center (NV$^-$) coupled to the spin of the nucleus of a 15-nitrogen atom and show that high fidelity gate operations are possible without the need for complicated composite pulse sequences. These operations include both the electron and nuclear spin rotations, as well as an entangling gate between them. These are experimentally realizable gates with current technology of sufficiently high fidelities that they can be used to build graph states for quantum information processing tasks. 
\end{abstract}
\maketitle

\section{introduction\label{intro}}
The quest to build quantum repeaters and computers, and to do communication with quantum processes, has been one of the most ambitious and difficult technological challenges of the 21st century so far. There have been many physical systems identified as potential candidates \cite{Ladd2010}. One that has enjoyed significant recent attention has been diamond. Diamond has many exceptional properties \cite{Coe2000,Lux1996,Zaitsev2001}; it is the hardest known material, chemically inert, possesses a broad optical transparency window, and can accommodate a large variety of optically active color centers. Of these color centers, the negatively charged nitrogen vacancy (NV$^-$) centre \cite{Davies1976,Hartley1984,Reddy1987}, has attracted particular interest \cite{Greentree2006}. It exhibits properties that make it useful for a wide range of interesting applications, including sensitive probes of magnetic fields \cite{Balasubramanian2008,Maze2008,Balasubramanian2009,Acosta2009} and biomarking tracking \cite{Chung2006,Fu2007,Chao2007,Neugart2007}. It also offers a quantum mechanical system that is remarkably isolated from the environment, ``trapped'' in a carbon lattice giving it excellent potential for quantum information processing based applications \cite{Beveratos2002,Beveratos2002a,Jelezko2004,Fuchs2011}.  

The NV$^-$ center is composed of an electron spin (spin $1$) and at least one nuclear spin coupled together by the hyperfine interaction \cite{Davies1976,Hartley1984,Reddy1987}. It is straightforward to manipulate both the NV$^-$ electron and nuclear degrees of freedom \cite{Smeltzer2009,Dutt2007,Jiang2009,Neumann2010,Robledo2011} and the long lived nuclear spin \cite{Childress2006} makes it attractive as a quantum memory \cite{Fuchs2011}. Furthermore, previous studies have shown that remote centers may be coupled using light \cite{Bernien2012}. Together, these features give us a tantalizing indication of a physical system to build quantum information processors with, as long as sufficient high precision operations are possible \cite{Childress2005,Childress2006a,Benjamin2006,Jiang2007,Barrett2005,vanderSar2013}. 

In this article we investigate the potential of a single NV$^-$ center to act with high fidelity as a node in a hybrid distributed quantum computer \cite{Childress2005,Childress2006a,Benjamin2006,Jiang2007,Barrett2005}. Our considerations will focus on the NV$^-$ center's electron spin with a coupling via the hyperfine interaction to a single \N{15} nuclear spin. The choice of \N{15} is motivated by the simpler nuclear energy level structure (spin $-\frac{1}{2}$), which is a natural qubit, compared with that of \N{14} (spin $1$). We will show that it is possible to enact a simple high fidelity universal set of quantum logic gates, with error probabilities  below the fault tolerance thresholds ($\sim$0.6\% \cite{Raussendorf2006}), between the electron and nuclear spins with a static magnetic field \cite{footnote1}. These gates consist of a hyperfine derived controlled phase gate, driven polarized microwave single qubit rotations on the NV$^-$ center and driven hyperfine mediated single qubit rotations on the nuclear spin. 

The paper is structured as follows: We begin in Section \ref{The computational basis} by defining our basic Hamiltonian process and the adoption of a rotating frame. We also discuss the critical physical parameters such as the relaxation and dephasing times of the electron and nuclear spins. Section \ref{The entangling gate} introduces the natural entangling gate between the electron and nuclear spin, while Sections \ref{Defect rotations} and \ref{nuclear rotations} show how electron and nuclear spins rotations can be achieved respectively with extremely high fidelity. Section \ref{circuits} shows how these basic operations can be used to build larger quantum circuits while Section \ref{A_graph_of_nuclei} presents a concluding discussion.

\section{The system Hamiltonian\label{The computational basis}}

\begin{figure}[htb]
\centering
\includegraphics[scale=1.2]{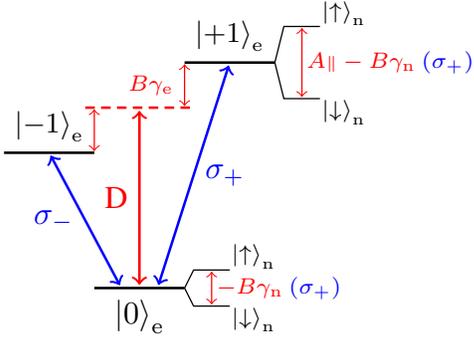}
\caption{[color online] Schematic energy level diagram of the ground state energy manifold of an NV$^-$ center. By setting a magnetic field such that $B\egyro<D$, a transition for the electron of the same circular polarization as that of the nucleus is chosen. The levels \kete{0} and \kete{+1} of the electron are chosen for the electron qubit (with the obvious mapping), and for the nucleus $\nup\mapsto\ket{1}$ and $\ndn\mapsto\ket{0}$.}
\label{driven-levels}
\end{figure} 

Our system under consideration here consists of a single $^{15}$NV$^-$ center whose $^3A$ ground state energy levels we illustrate in  Figure \ref{driven-levels}. The NV$^-$ is made up of an electron spin 1 system and a nitrogen-15 nuclear spin. As electron spin consists of three levels we must choose a pair of levels to embed the electron qubit in ($\kete{0},\kete{+1}$  or $\kete{0},\kete{-1}$). The nuclear spin from the nitrogen is naturally a spin $-\frac{1}{2}$ system. At this stage we are going to describe our system in terms of the Hamiltonian $H = H_0 + H_\mathrm{S} + H_\mathrm{HF} + H_\mathrm{D}$ where 
\begin{align}\label{Schrodinger_Hamiltonian}
H_0 &= \hbar DS_z^2+  \hbar B\egyro S_z-  \hbar B\ngyro I_z\,,
\end{align}
\begin{align}
H_\mathrm{S} &=  \hbar \frac{E}{2}\left(S_+^2 + S_-^2\right)\,,\\
H_\mathrm{HF} &= \hbar \Apara S_zI_z+ \tfrac12  \hbar \Aperp\left(S_+I_- + S_-I_+\right)\,,\\
H_\mathrm{D} &= \hbar  \Omega_0\cos\left(\omega t+\phi\right)\left(S_x - \frac\ngyro\egyro I_x\right)\,,
\end{align}
with $S_{x,y,z}$ being the usual spin $1$ operators ($S_+=\sqrt{2}\left(\kete{0}\brae{-1}+\kete{+1}\brae{0}\right)$, $S_-=S_+^\dagger$) and $I_{x,y,z}$ the nuclear spin $-\frac{1}{2}$ operators ($I_+=\nup\bran{\downarrow}$, $I_-=\ndn\bran{\uparrow}$).  The uncoupled system is given by the Hamiltonian terms $H_0+H_\mathrm{S}$ where the first term in $H_0$ is a zero-field splitting of magnitude $D/2\pi=2.87\,\text{GHz}$ \cite{Felton2009}. The second term  is the splitting of the three electron levels determined by the magnetic field $B$ aligned with the NV-axis and the electron gyromagnetic ratio $\egyro = \mu_\text{B}g_\text{e}$ where $\mu_\mathrm{B}/h=14.0\,\text{MHz\,mT}^{-1}$ and $g_\mathrm{e}=2.00$. The last term in $H_0$ is the splitting of the two level nuclear levels determined by the magnetic field $B$ and the nuclear gyromagnetic ratio $\ngyro = \mu_\mathrm{N}g_\mathrm{n}$ with the nuclear magneton $\mu_\mathrm{N}/h=7.63\,\text{kHz\,mT}^{-1}$ and the nuclear $g$-factor $g_\mathrm{n}=-0.566$. The second term in our  uncoupled system $H_\mathrm{S}$ represents a strain induced splitting between the $\kete{\pm1}$, where $E\sim 1 - 10\,\text{MHz}$ is typical (7 MHz is assumed here). The hyperfine coupling $H_\mathrm{HF}$ between the electron and nuclear spins is composed of a phase gate with $\Apara/2\pi=3.03\,\text{MHz}$ parallel to the NV$^-$  axis, and an exchange part with $\Aperp/2\pi=3.65\,\text{MHz}$ perpendicular to it. The exchange component of  hyperfine interaction moves an excitation between the electron and nuclear spin systems and has its resonance at $B_\mathrm{ex} = \frac{\Apara/2\mp D}{\egyro\pm\ngyro} \approx\mp102\,\text{mT} $  with Lorentzian width $B_\mathrm{FWHM} = \frac{2\sqrt{2}\,\Aperp}{\ngyro+\egyro} \approx0.368\,\text{mT}$ \cite{exchange}. Finally, $H_\mathrm{D}$ is an electromagnetic driving term, with a magnitude on the electron (nucleus) determined by the gyromagnetic ratio $\egyro$ ($\ngyro$), and $\Omega_0$ is the amplitude of the applied electromagnetic drive field of frequency $\omega$ and phase $\phi$.

Before moving forward to examine various gate operations we need to examine the physical and coherence properties of our system. Beginning with the electron spin, the type of synthesis used to create the diamond crystal and temperature have a significant effect on the relaxation time $T_1$ time of the electron spin. At temperatures $T > 200$K, Jarmola et.al \cite{Jarmola2012} reported that high-pressure, high-temperature (HPHT) and chemical vapor deposition (CVD) samples showed the same $T_1$ time within a factor of 2. However for lower temperatures the relaxation time can dramatically increase. For instance with CVD samples the $T_1$ time could increase by almost 5 orders of magnitude (to 100s) when the temperature is decreased below 80K. This seems to strongly indicate that we want to work with a CVD diamond at moderate temperatures (4-80K). Furthermore the relaxation time is also increased when a small but nonzero (20mT) external magnetic field is applied. Applying a magnetic field allows us to split the $\kete{\pm 1}$ levels. Next the dephasing time $T_2$ is also highly dependent on the type of diamond sample used and especially the concentration of P1 centers (and other impurities). Electronic spin coherence times $T_2^*$ of  $90\mu$s and  $T_2 > 1.8$ ms have been observed in isotopically purified diamond \cite{Maze2008,Ishikawa2012,Fang2013}. The coherences properties of the nuclear spins can exceed 1s \cite{Maurer2012}.

\section{The entangling gate\label{The entangling gate}}

Before we begin examining gate operations we need to specify how we are going to encode a qubit within the electron spin. In this case we are simply going to choose the electron spins computational basis \kete{0} and \kete{1} to be \kete{0} and \kete{+1} respectively. The \kete{-1} state should never be occupied and if it is, it will be considered a leakage event. Now the hyperfine interaction between the electron and nuclear spins provides a route to entangling the spins without resorting to driving fields or varying the magnetic fields dynamically. In the case of no driving field, the interaction picture Hamiltonian reduces to
\begin{eqnarray}
\bar{H} &= & \hbar \Apara S_zI_z+ \hbar \left(E \ket{+1}\bra{-1}\expup{i2B\egyro t} \right. \\
&+& \left. \frac{\Aperp}{\sqrt{2}} I_- \left( \expup{i\Delta_+t} \kete{+1}\brae{0} + \expup{-i \Delta_-t}  \kete{0}\brae{-1} \right)+\herm \right) \nonumber
\end{eqnarray}
where $\hbar \Apara S_zI_z$ is a natural entangling gate (a controlled phase gate) and $\Delta_\pm= D\pm B\egyro\pm B\ngyro$.  However  the strain induced splitting is obviously problematic, since it provides a transition out of the computational basis of the system. The resonances for this strain interaction occur when $B = \pm\frac{\Apara}{2\egyro}$ (positive for the nucleus is in the \ndn\ state, and negative for the \nup\ state) with full width at half maximum $B_\text{FWHM}=2E/\egyro$. For a strain induced splitting $E<10\,\text{MHz}$ the width is $B_\text{FWHM}<1\,\text{mT}$ approximately centered at 0\,T. Eliminating this leakage effect is as simple as choosing $B\gg E/\egyro$, which is certainly true for a choice of $B\sim 50\,\text{mT}$. In such a case the strain term gives a small dispersive interaction of the form $\bar{H}_\mathrm{S,dis} = \frac{ \hbar E^2}{2B\egyro} \left[ \kete{+1}\brae{+1} - \kete{-1}\brae{-1}\right]$. At the choice of $B\sim 50$ mT we are also far off-resonance with the exchange component of  hyperfine interaction and so its contribution is in terms of a small dispersive shift.

\begin{figure}[ht]
\centering
\includegraphics{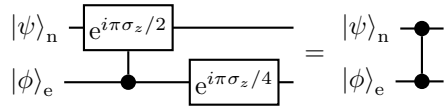} 
\caption{Quantum circuit showing the equivalence between the hyperfine interaction and a controlled phase gate. The interaction provided by the hyperfine Hamiltonian (the controlled gate on the left) is locally equivalent to a CZ gate. As the single qubit operation on the electron commutes with the  controlled gate, it may appear on either side of it.\label{CPhase-circuit}}
\end{figure}

In these regimes we will have no (or extremely small) exchange population between the electron spins states and so long as we start our electron spin in its qubit subspace $\{\kete{0},\kete{+1} \}$ we can neglect the $\kete{-1}$ state. This means we can write as entangling operation in the $\{\kete{0},\kete{+1} \}$ subspace as
\begin{eqnarray}\label{effham}
\bar{H} &= & \left( \hbar \Apara + \frac{ \hbar \Aperp^2}{2 \Delta_+}\right) \kete{+1}\brae{+1}  I_z  \\
&-& \left(\frac{ \hbar \Aperp^2}{2 \Delta_+} + \frac{ \hbar \Aperp^2}{2 \Delta_-}\right) \kete{0}\brae{0}  I_z  \nonumber \\
&+& \left( \frac{ \hbar E^2}{2B\egyro} -  \frac{ \hbar \Aperp^2}{4 \Delta_+}\right) \kete{+1}\brae{+1}  \nonumber \\
&-&  \left(\frac{ \hbar \Aperp^2}{4 \Delta_+} - \frac{ \hbar \Aperp^2}{4 \Delta_-}\right) \kete{0}\brae{0}  \nonumber
\end{eqnarray}
which includes level shifts \cite{Shore1981} due to the strain induced splitting and the perpendicular hyperfine interaction. To lowest order, this gives an interaction of the form
\begin{equation}
\bar{H} \approx \frac{ \hbar  \Apara}{2}  \kete{1} \brae{1} \sigma^\mathrm{n}_z\,,
\end{equation}
The interaction is equivalent to a controlled phase gate within a single qubit $z$-rotation on the electron, as shown in Figure \ref{CPhase-circuit}. 
For instance an ideal input state to create a maximally entangled pair is \nvket{+}{+}. The gate time is determined by $\Apara$, which gives a time of $t_\mathrm{CZ}=\pi/\Apara\approx165\,\text{ns}$. However looking Equation \eqref{effham}, we notice there are second order terms which could be important (and will affect the fidelity of this gate operation). Hence we perform a simple simulation using the full Hamiltonian \eqref{Schrodinger_Hamiltonian} but with the driving turned off. This is depicted in Figure {\ref{CPhasesim} and clearly shows that an extremely high fidelity gate in principle can be achieved ($<0.0001$). The fidelity is limited not in this case by the decoherence properties of the NV$^-$ centers (more explicitly $T_2^*$ of the electron spin) but by the strain and $\Aperp$ components of the hyperfine interaction. These effects could in principle be corrected by appropriate single qubits operations on the electron and nuclear spins. We did not do this in the simulations as we wanted to keep the gate requirements are simple as possible. Even without such corrections, a high fidelity gate is possible when one has a long $T_2^*$ for the electron spin. Finally our simulations indicate for this gate that the leakage rate to the \kete{-1} state is exceedingly small. 
\begin{figure}[ht]
\centering
\includegraphics[scale=0.9]{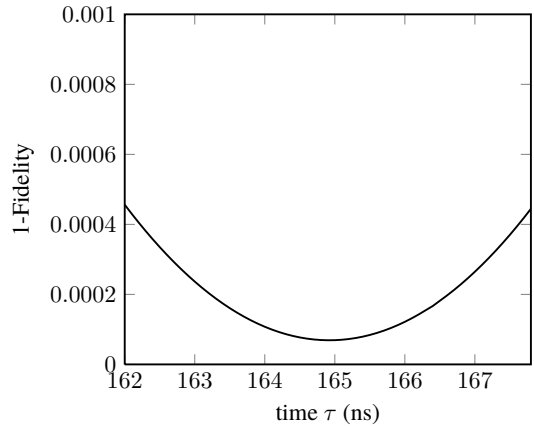} 
\caption{Simulation illustrating the probability of error (1 - Fidelity) of the CZ gate when applied to an initial electron nuclear spin state \nvket{+}{+} to generate a maximally entangled Bell state..\label{CPhasesim}}
\end{figure}

\section{Electron rotations\label{Defect rotations}}

\begin{figure*}[htb]
\centering
\includegraphics[scale=0.65]{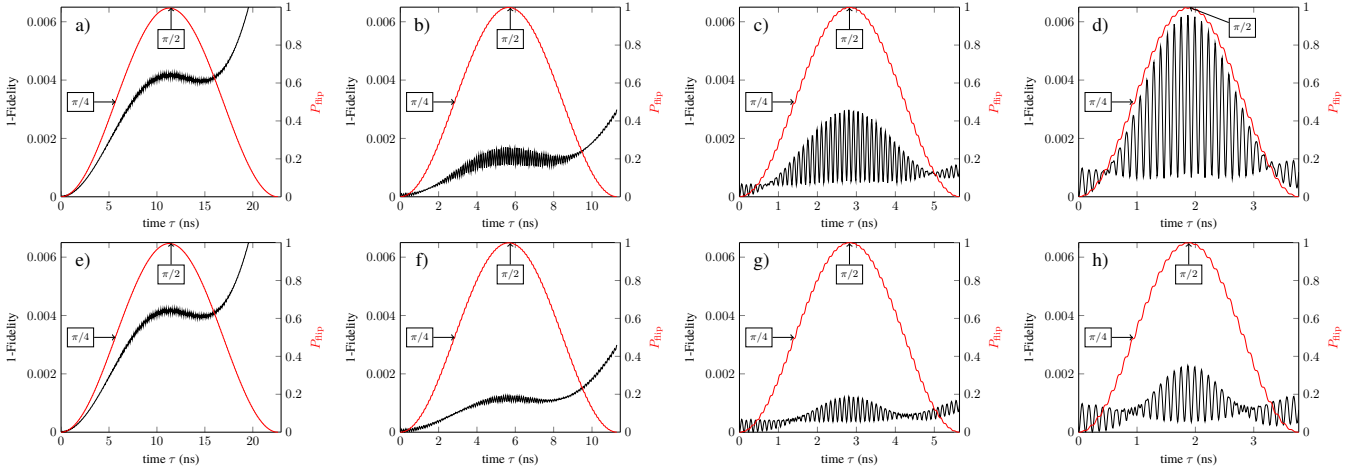}
\caption{Simulation illustrating the probability of error (1 - fidelity) of the driven system compared to the ideal electron spin qubit rotations versus rotation time. An unpolarized microwave field is used in (a-d) while a polarized field is used in (e-h). Four different drive amplitudes were considered, $62.5\,\text{MHz}$ for (a \& e), $125\,\text{MHz}$ for (b \& f), $250\,\text{MHz}$ for (c \& g) and $375\,\text{MHz}$ for (d \& h). The black curves show the probability of error of the gate operation while the red curves show the degree of rotation. The rotation angle corresponds to the definition of the gate $R_y(\theta)=\cos(\theta/2)\openone+i\sin(\theta/2)\sigma_y$. The $\pi/4$ and $\pi/2$ points are explicitly indicated. For these simulations, a full master equation was used with appropriate  relaxation and dephasing rates for both the electron and nuclear spins (detailed in the appendix).  The counter rotating terms cause the very rapid small oscillations. 
\label{model-electron}}
\end{figure*}

Rotations on the electron spin can be simply implemented using a microwave driving field perpendicular to the NV$^-$  axis without the same drive also affecting the nuclear spin. This assumption is safe, since the gyromagnetic ratio of the electron is $\sim6500$ larger than that of the nucleus and the microwave driving field is far off-resonance with it. However, we need to be careful that the field used to drive the $\kete{0} \leftrightarrow \kete{+1}$ transition does not also drive the $\kete{0} \leftrightarrow \kete{-1}$ transition.  This is true when 
\begin{equation}\label{electron-one-transition}
\frac{\Omega_0}{2\sqrt{2}}\ll \min\{2B\egyro,2D\}\,,
\end{equation}
which is equivalent to the statement that the coupling factor of the drive field to the electron is much less than the detuning between the transition that we want to drive and the transition that we don't want to drive. The limit has been demonstrated experimentally \cite{Fuchs2009}, and becomes apparent at attempted cycle times of a few tens of nanoseconds. In Figure \ref{model-electron}(a-d) we show the error probability (1-fidelity) for a single qubit rotation  achieved in less than $10\,\text{ns}$ using four different pump strengths ($62.5\,\text{MHz}$, $125\,\text{MHz}$, $250\,\text{MHz}$ and $375\,\text{MHz}$). At $62.5\,\text{MHz}$ the hyperfine interaction is the primary cause of the error probability as we have assumed that the nuclear spin is in an equal superposition state. Still for a $\pi/4$ rotation, the probability of error is approximately 0.0035. As  we increase the driving field strength the error probability becomes dominated by the breakdown of the rotating wave approximation (the counter rotating terms) and weak driving of the $\kete{0} \leftrightarrow \kete{-1}$ transition. The optimal working point (that minimizes both error sources) is a driving field around 125 MHz where an error probability of $\sim 0.002$ is possible. For many quantum information processing tasks (computation or communication) this may not be low enough. 

There are a number of simple ways to decrease this error probability. Remember that for the fastest gates we were beginning to populate the $\kete{-1}$ state. We can use a polarized drive field such that only the $\kete{0} \leftrightarrow \kete{+1}$ transitions is selected. This occurs for  magnetic fields $B\egyro<D$. We plot the results of our simulations in Figure \ref{model-electron}(e-h). In these cases we are looking at a simple two level driven system which. It is still subject to the rotating wave approximation and can break down if the driving amplitude is larger than the transition frequency. With a rotating wave approximation, $\nu=D+B\egyro$ and $\phi=-\pi$ the dynamics is governed by an effective Hamiltonian of the form
\begin{equation}\label{HapproxS}
H \approx \hbar  \Apara \kete{+1} \brae{+1} I_z -  \frac{\hbar \Omega_0}{2\sqrt{2}}\big(\kete{0}\!\brae{+1} + \kete{+1}\!\brae{0}\big)\,,
\end{equation}
from which we can immediately confirm the intuition that the electron gates must operate much faster than the entangling gate time to avoid the hyperfine shift. That is,
\begin{equation}\label{electron-drive-faster}
\Apara \ll \frac{\Omega_0}{\sqrt2}\,.
\end{equation}
The conditions imposed by \eqref{electron-one-transition} and \eqref{electron-drive-faster} imply an optimal drive frequency, which for a magnetic field of 50\,mT is approximately $\frac{\Omega_0}{2 \pi} \sim 250\,\text{MHz}$ as found by our numerical study. In such a case the $x$-rotation gate
\begin{equation}
R_\mathrm{e}^x=\exp\left(i\frac{\Omega_0}{2\sqrt{2}}  t \big[\kete{0}\!\brae{+1} + \kete{+1}\!\brae{0}\big]\right)\,
\end{equation}
follows trivially. Similarly, the $y$-rotation gate is produced when the $\phi=-\pi/2$ is chosen. A comparison of the model with a numerical simulation of the system with appropriate decay probabilitys is shown in Figure \ref{model-electron}(e-h). With $\frac{\Omega_0}{2 \pi} \sim 250\,\text{MHz}$ we can achieve a $\pi/4$ ($\pi/2$) rotation in approximately $1.5\,\text{ns}$ ($3\,\text{ns}$) with an error probability less than 0.0005 (0.001). These error probabilitys are within what is typically required for large scale quantum computation. The main cause of the loss in fidelity is associated with the $\Apara$ hyperfine interaction and  the breakdown of the rotating wave approximation. If one requires even higher fidelity gates, one can use composite pulses to decouple the nuclear spin, but at the cost of slower gates.

\section{Nuclear Rotations\label{nuclear rotations}}

The use of the nuclear spins either as a memory or as computational qubits has both advantages and disadvantages. First the \N{15}\ nuclear spin is spin half and so we do not need to worry about populating other energy levels when we drive it. Also the comparative weakness of the nuclear gyromagnetic ratio provides isolation of the nucleus from electron rotations regardless of polarization. However, this comes at the cost of difficulty in affecting rotations on the nuclear spin qubit without also affecting the electron spin. To avoid driving the electron spin the time needed for nuclear gates will be much longer than for the entangling operation.  However  this longer time means the hyperfine interaction cannot be neglected during nuclear spin rotations. The  consequence of this is that the hyperfine shift leads to a driving frequency for the nucleus conditioned on the state of the electron, and that nuclear gates will need to be clocked on the entangling gate. This could also  be an issue in other model systems for quantum computation. We can largely avoid this problem because nuclear rotations are only necessary upon initialization and measurement, which are both steps for which the electron is in a separable state and therefore can be manipulated with impunity. 

\begin{figure}[htb]
\centering
\includegraphics[scale =0.9]{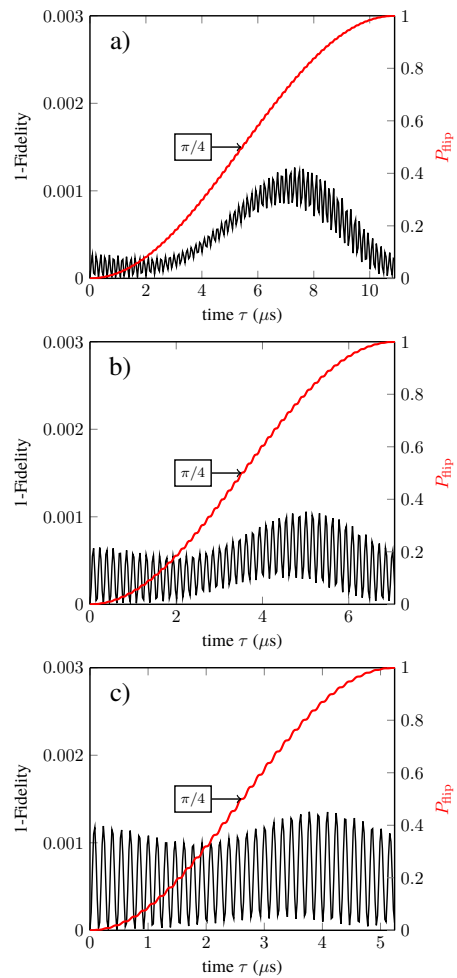}
\caption{Simulation showing the fidelity for nuclear spin rotations under both varying driving field and electron and nuclear spin decoherence. The nuclear gate takes the longest to perform, and is thus most prone to error. We show three different driving field strengths: 91 MHz a), 140 MHz b) and 189 MHz c). The black curves show the probability of error of the gate operation while the red curves show the degree of rotation. The rotation angle corresponds to the definition of the gate of the form $R_y(\theta)=\cos(\theta/2)\openone+i\sin(\theta/2)\sigma_y$. The $\pi/4$ points are explicitly indicated. For these simulations, a full master equation was used with appropriate  relaxation and dephasing rates for both the electron and nuclear spins (detailed in the appendix).  
\label{nuclear-fidelity}}
\end{figure}

The obvious choice of the electron state is the polarized ground state \kete{0} where the electron and nuclear spins are effectively decoupled. The parallel hyperfine interaction imparts no additional phase between the two subsystems. In other words, whilst the frequency of the nuclear drive field is conditioned on the state of the electron ($\Apara\gg\Omega_0\ngyro/\egyro$), we never need to consider this as the electron state is always the same. This argument however neglects thermalization effects on the electron spin which occurs at time scale around 1 s (for room temperature) or longer (for lower temperatures). With operation times around 50 $\mu$s it could be difficult to achieve very high fidelity operations. 

In the $B\egyro<D$ regime, the Zeeman splitting of the nucleus is an order of magnitude smaller than the hyperfine splitting, which contributes to the $\nvket{+1}{\downarrow}\leftrightharpoons\nvket{+1}{\uparrow}$ transition, but not the $\nvket{0}{\downarrow}\leftrightharpoons\nvket{0}{\uparrow}$ transition. Hence by initializing the electron spin  in \kete{+1}, a larger drive field can be used before the rotating wave approximation on the nuclear rotation breaks down, implying a faster nuclear rotation operation. For the nuclear spin, two processes are involved in the drive between the \nup\ and \ndn\ states. The obvious is the direct drive shown in $H_\mathrm{D}$. The other process is second order, involving the off resonant drive on the electron and the perpendicular hyperfine term. As the gyromagnetic ratio of the electron is large compared with that of the nucleus, this second order term actually dominates at the magnetic field we have selected, which is necessary for useful nuclear gate times. Operating in a regime where the state of the electron is unchanged by the drive field of frequency
\begin{equation}
\nu = \Apara-B\ngyro+\frac{\Aperp^2}{2D+2B\egyro}\,,
\end{equation}
and where the perpendicular hyperfine term is off resonance, it is straightforward to show in our rotating frame that we perform a single qubit nuclear rotation of the form
\begin{equation}
R_\mathrm{n}^\phi=\exp\left(-i \frac{\Omega_0}{4}\left|\frac{\Aperp}{D+B\egyro}-\frac\ngyro\egyro\right|  \sigma^\mathrm{n}_\phi t\right)\,,
\end{equation}
where $ \sigma^\mathrm{n}_0 =  \sigma^\mathrm{n}_x$ and  $\sigma^\mathrm{n}_{-\pi/2}= \sigma^\mathrm{n}_y$. The two terms in  $R_\mathrm{n}^\phi$ can be opposing such that for a magnetic field of $B=-0.9477\,\text{T}$ no driving of the nuclear spin is possible due to destructive interference. As were are working in the regimes $0\leq B\egyro  \leq D$ our gate is fastest for $B\egyro$ as small as possible. Now as the nuclear gate takes a relatively long time to perform, it is the most sensitive to error. Figure \ref{nuclear-fidelity} shows the results of a simulation of the nuclear spin rotation for various driving fields when the electron spin is prepared in the \kete{+1} state. For a driving field $\frac{\Omega_0}{2 \pi} = 140\,\text{MHz}$, the gate time required to rotate the nuclear spin from  \ketn{0} to \ketn{+} takes 3.54\,\textmu s with a error probability less than 0.001  as is shown in Figure \ref{nuclear-fidelity}(b). This  can be compared with a gate time of 57.8\,\textmu s for a similar rotation (but a slightly worse error probability of 0.003) if the electron spin was polarized in the \kete{0} state. Our 0.001 error estimate is also on the high side as we are taking the maximum value of the oscillating curve at a given time rather than its minimum. This is very conservative. The error probability can be dramatic reduced to below 0.0003  (see Figure \ref{nuclear-fidelityideal}) just by choosing our gate time accurately to work at the minimum. In such a case our timing accuracy needs to be below 10ns. By increasing the strength of the driving field, we can in principle achieve a $1\,\text{\textmu\,s}$ $\pi/4$ rotation with error probability below 0.001.

\begin{figure}[htb]
\centering
\includegraphics[scale =0.2]{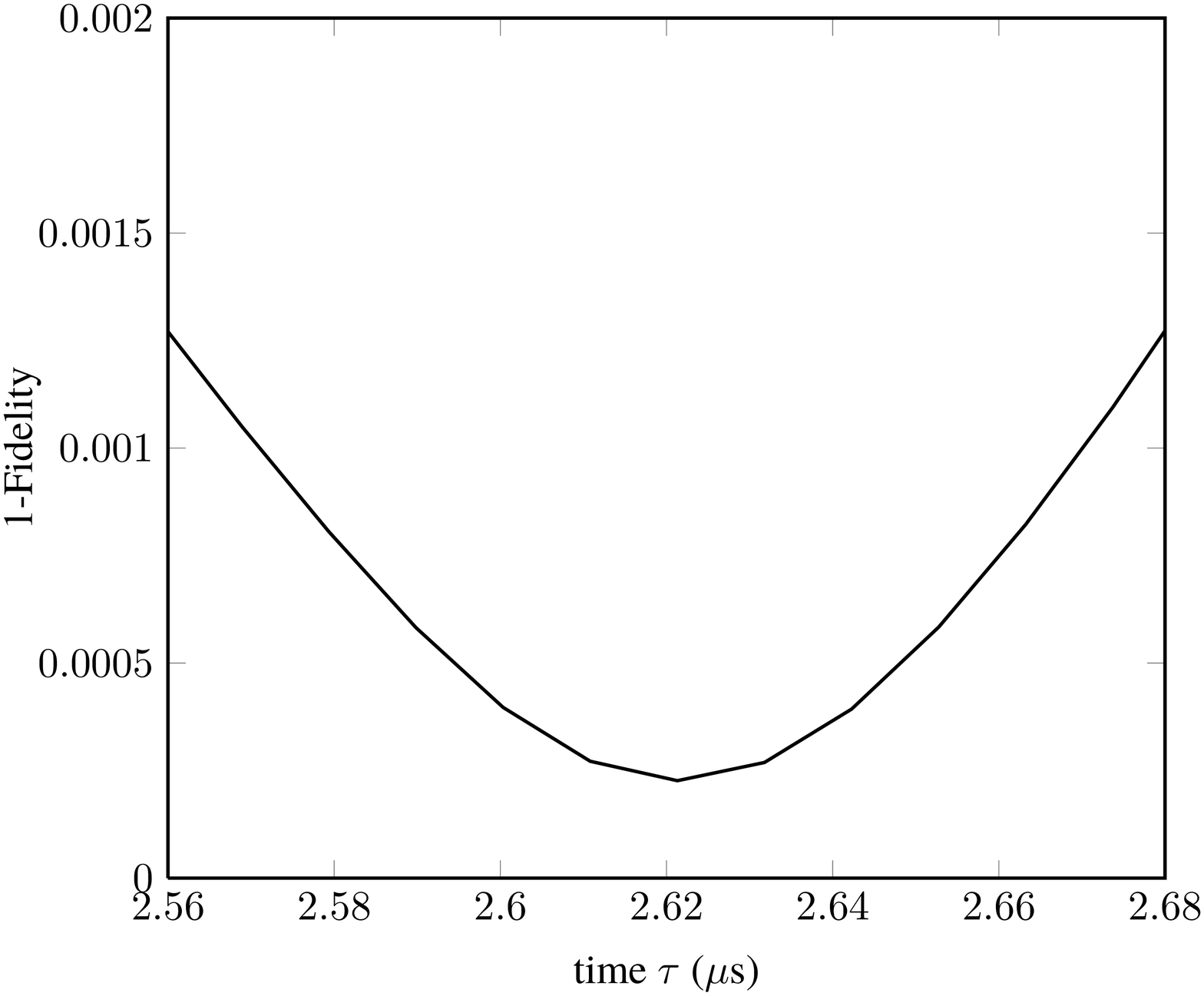}
\caption{Simulation showing the error probability ($1-\bran{+} \rho (t) \ketn{+}$) for nuclear spin rotations under a 189 MHz driving field for a $\pi/4$ $y$ rotation versus time.  
\label{nuclear-fidelityideal}}
\end{figure}

The results for both the electron and nuclear spin rotations as well as the hyperfine mediated CZ gate show such operations can be achieved under realistic conditions with error probabilities of 0.001 or less \cite{footnoten}. Such error probabilitys are generally what are going to be required for quantum information processing tasks. Generally to achieve the faster nuclear spin  rotations we needed to polarize the electron spin in the \kete{+1} state. For arbitrary quantum computation this may not be convenient, however it works well for cluster state techniques. In such situations we begin by initializing the nuclear spin in the  \ketn{+} state while the electron spin is in the \kete{+1} state. An arbitrary state can then be created in the electron spin (or between various electron spins via optical channels for instance). A CZ gate between the electron and nuclear spins via the hyperfine interaction can then be used to entangle them. A simple $\pi/4$ rotation of the electron spin followed by its measurement teleports the state of the electron spin to the long lived nuclear spin. In principle this allows a cluster state to be generated between various nuclear spins \cite{Nemoto2013}.

\section{Circuits and implementation}\label{circuits}

\begin{figure}[htb]
\includegraphics{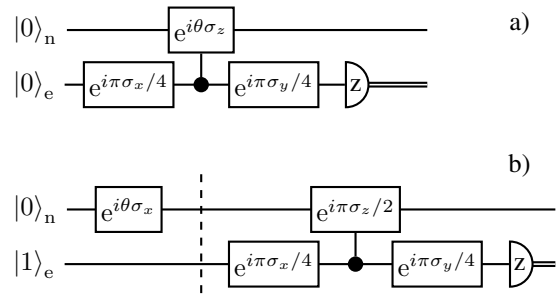}
\caption{Characterization of the hyperfine interaction. a) By varying the time that the hyperfine interaction is allowed to operate for (the waiting time between $x$ and $y$ rotations on the electron) the probability of measuring the electron in \kete{0} is $P_{\ket{0}} = \sin^2(\theta/2)$. b) Characterization of the nuclear spin. By varying the input state of the nuclear spin and with a set waiting time between electron rotations, the probability of finding the electron in the \kete{0} state is $P_{\ket{0}}=\sin^2(\theta)$. The dashed line is placed to emphasize that the nuclear rotation is performed only when the electron is in the \kete{1} state, so that the splitting between the \nup\ and \ndn\ states is enhanced by the hyperfine interaction (which is much larger than the magnetic splitting alone). This circuit can be employed to measure the state of the nucleus in the $z$ basis.
\label{zz-character-circ}}
\end{figure}

Before implementing quantum circuits, the NV$^-$  center must be properly characterized. This can be done in a three tiered way, starting with characterizing the electron spin, followed by the controlled phase interaction, and finally the nuclear spin. The case of the electron is nearly trivial, and simple Rabi nutation can be measured optically \cite{Fuchs2009} provided that the nucleus can be polarized to avoid interference from the hyperfine shift. Ultimately, once the characterization of a center is complete, the polarization of the nucleus should be done using a circuit to effectively swap its state with that of a polarized electron. However, for characterization purposes the exchange component of the hyperfine interaction can be use to do this by preparing the electron in a polarized state and then sweeping the field over the exchange resonance. Circuits for characterizing the controlled phase interaction and nuclear rotations are given in Figure \ref{zz-character-circ}.

With a fully characterized system, the repetitive nature of measurement based quantum computers means that only a few non-trivial quantum circuits are needed. Specifically, these are initialization of the nucleus, entangling of the electron and nucleus, and measuring of the nucleus via the electron. For initialization of the nucleus, it is clear that we must know the state of the nucleus or be able to project the nuclear spin into a known state, which relates it to the measurement of the nuclear spin. In fact, the gates needed are simple applications of the two characterization circuits in Figures \ref{zz-character-circ} (characterization and use of electron rotations is considered trivial). By fixing the nuclear rotation angle and axis in the circuit in Figure \ref{zz-character-circ}  the nuclear state can be measured in a particular basis. Application of this circuit, followed by a subsequent reset of the electron state into \kete{1} and a nuclear rotation qualifies as a preparation circuit for the nucleus. Application of the entangling gate consists of no more than a waiting time between single qubit operations. With these operations and a probabilistic mechanism to entangle remote electron spins we can then undertake various quantum communication and distributed quantum computation tasks.

\section{Concluding discussion\label{A_graph_of_nuclei}}

In this paper we have shown a universal set of operations which can be implemented in a NV$^-$ center with a nitrogen-15 nucleus.  The fastest operation is the electron spin rotation, while the medium-term operation is the entangling gate  which uses the hyperfine interaction. The electron spin rotations can be done on timescale such that the nuclear spin is not affected.   For the entangling gate, we utilize the controlled-phase interaction component rather than the exchange interaction component of the hyperfine coupling.  By sweeping the magnetic field, the exchange interaction can be turned on and off making it useful for quantum information processing.  However, the controlled-phase interaction is always on and hence during the iSWAP operation there will be a contribution from the phase part which will need to be corrected.  In addition, any imperfection in the iSWAP operation will lead to depopulation of the nuclear spin, an error channel, which will shorten the effective coherence time. Instead, a static field can be used that is far detuned from the exchange interaction resonance to implement  a highly efficient controlled-phase gate.  This is preferable in terms of errors and stability coming from the static magnetic field.  Finally, the longest time operation is the nuclear spin gate.  This can be done by driving the nuclear spin weakly enough that we avoid disturbing the electron.  In the case of distributed quantum computation, the nuclear spin rotation is usually required only twice, first at the beginning and then just before its measurement.  Hence the time necessary for this gate is not crucial as the other gates. However, considering the potentially probabilistic nature of coupling two remote electron spins, it would be better to have as quick an operation as possible.  Our numerical simulations show that we can achieve a error probability lower than that which the thresholds indicate is required for distributed quantum computer.  These operations are done without the need for complicated composite sequences which should make the complexity in building larger scale NV$^-$ center based devices much more manageable. It is now critical to re-emphasize that our results are based on operating temperatures between 4-80K (and not room temperature). This is because we want our electron spin $T_1$ relaxation time approximately $10^4$ time greater than our longest gate operation (the nuclear spin rotation) which takes a few microseconds to perform. It may be possible to engineer samples that operate at room temperature with this coherence properties. Finally while the focus of this paper has been on NV$^-$ centers with a nitrogen-15 nucleus, our results can be applied to many other solid state systems.

\begin{acknowledgments}
We acknowledge support from JSPS, FIRST, and NICT in Japan.
\end{acknowledgments}

\appendix

\section{The interaction Picture}

The Hamiltonian for the electron spin - nuclear spin NV$^-$ center hybrid system can be written as
\begin{align}\label{Schrodinger_Hamiltonian1}
H &= H_0 + H_\mathrm{S} + H_\mathrm{HF} + H_\mathrm{D}\,,
\intertext{where}
H_0 &= \hbar DS_z^2+  \hbar B\egyro S_z-  \hbar B\ngyro I_z\,,\\
H_\mathrm{S} &=  \hbar \frac{E}{2}\left(S_+^2 + S_-^2\right)\,,\\
H_\mathrm{HF} &= \hbar \Apara S_zI_z+ \tfrac12  \hbar \Aperp\left(S_+I_- + S_-I_+\right)\,,\\
H_\mathrm{D} &= \hbar  \Omega_0\cos\left(\omega t+\phi\right)\left(S_x - \frac\ngyro\egyro I_x\right)\,,
\end{align}

For our purposes, it is convenient to move to an interaction picture defined by $\bar{H}=\expup{iH_0t/\hbar }H\expup{-iH_0t/\hbar}-H_0$. In such a situation Equation \eqref{Schrodinger_Hamiltonian} simplifies to  $\bar{H}_{\rm eff}= \bar{H}_\mathrm{S}+\bar{H}_\mathrm{HF}+\bar{H}_\mathrm{D}$, where
\begin{eqnarray}
\bar{H}_\mathrm{S}& =& \hbar \ E \kete{+1}\brae{-1}\expup{i2B\egyro t} +\herm\,, \nonumber  \\ 
\bar{H}_\mathrm{HF} &=&\hbar \ \Apara S_zI_z+\hbar  \frac{\Aperp}{\sqrt{2}}\bigg[I_- \kete{+1}\brae{0}\expup{i\Delta_+ t} \nonumber \\
&&\hspace{0.2cm}+I_-\kete{0}\brae{-1}\expup{-i \Delta_- t}+\herm\bigg]\, \nonumber \\
\bar{H}_\mathrm{D} &=&   \frac{\hbar \Omega_0 }{\sqrt{2}}\cos\left(\nu t+\phi\right)\bigg[\kete{+1}\brae{0}\expup{i(D+B\egyro)t} \nonumber\\
&&\hspace{0.05cm}+\kete{0}\brae{-1}\expup{-i(D-B\egyro)t}+\frac{\ngyro}{\sqrt {2} \egyro} I_+\expup{-iB\ngyro t} \bigg]+\herm \nonumber
\end{eqnarray}
with $\Delta_\pm= D\pm B\egyro\pm B\ngyro$. The parallel term in $\bar H_\mathrm{HF}$ remains unchanged in this interaction picture, since it commutes with $H_0$. 


\section{Resonances and the strain induced splitting}
The strain induced splitting from our Hamiltonian causes population to oscillate between the \kete{+1} and \kete{-1} electron spin states. If the energy difference between these states is quite small, then population may move completely from one state to the other. In our system the \kete{-1} state is outside of the computational basis, and so we want to avoid the action of strain induced splitting. We can derive the width and position of this resonances with respect to the applied magnetic field along the NV$^-$  axis. Starting with Equation \eqref{Schrodinger_Hamiltonian} we will neglect the driving term and the perpendicular hyperfine term (the perpendicular hyperfine term is far  from resonance with the low magnetic fields considered here). In addition, the strain induced splitting does not involve the \kete{0} state, so we may disregard this state for the moment. We are this left with the reduced Hamiltonian
\begin{equation}
H = \hbar B\egyro S_z -  \hbar B\ngyro I_z +  \hbar A_\parallel S_zI_z + \hbar  \frac{E}{2}(S_+S_+ + S_-S_-)\,. \nonumber
\end{equation}
There are two resonances, which are dependent on the state of the nucleus through the parallel hyperfine term and the nuclear magnetic term.
\begin{align}
H_{\nup} &=  \hbar B\egyro S_z + \hbar \frac{A_\parallel}{2}S_z +  \hbar\frac{E}{2}(S_+S_+ + S_-S_-)\,, \nonumber \\
H_{\ndn} &=  \hbar B\egyro S_z - \hbar \frac{A_\parallel}{2}S_z + \hbar \frac{E}{2}(S_+S_+ + S_-S_-)\,. \nonumber
\end{align}
Selecting the nucleus in the \nup\ state the resultant $2\times2$ Hamiltonian has equal diagonal terms when $B=-\Apara/2\egyro$ and $B=+\Apara/2\egyro$ when the nucleus is in the \ndn\ state, giving us the resonances at $\mp 0.054\,\text{mT}$.  Taking the hyperfine terms into account leads to a shift of these resonances by $\approx 4.6\,\text{\textmu T}$. Again assuming that the perpendicular hyperfine interaction may safely be neglected the resonances have Lorentzian profiles of $B_\text{FWHM}=2E/\egyro$. For a strain induced splitting of the order $E\sim 1\,\text{MHz}$ our chosen magnetic field of $B=50\,\text{mT}$ is safely distant from these resonances.

\section{Resonances and the hyperfine coupling}

When a electron transition is close in frequency to the nuclear transition, an excitation may oscillate between them. In isolation, the perpendicular hyperfine interaction provides an iSWAP gate, which is entangling. However, the exchange interaction is not in isolation due to the parallel hyperfine term in the Hamiltonian, which may not be tuned to be negligible. In addition, tuning the exchange interaction increases the complexity of the gate, as it would require the fine control of a strong magnetic field. An added incentive to avoid an exchange interaction is the impact that errors in the interaction would have on the large scale system. This interaction was recently demonstrated as a quantum memory \cite{Fuchs2011}, for which the nucleus is initially polarized and iSWAP and SWAP are the same up to a global phase.  It is conceivable that the inverse operation could be used to initialize the nuclear state by tuning the exchange interaction into resonance and polarizing the electron \cite{Smeltzer2009}. This offers a potential alternative to the initialization and readout operations that we describe in Section \ref{circuits}.


\section{Master equation}

We need to be able to model the various electron spin, nuclear spin rotations as well the controlled Z gate. We need to able to include the effects of decoherence (thermalization) and dephasing on both the electron and nuclear spins. These can be achieved using the master equation
\begin{eqnarray}
\frac{\partial\bar\rho}{\partial t} &=& -\frac{i}{\hbar}\left[\bar H_\text{eff},\bar\rho\right] +\Gamma_n^{(2)}\left(\sigma_z \bar\rho \sigma_z  - \bar\rho \right) \nonumber \\
&+& \frac{\Gamma_e^{(1)} (\bar n_e+1) }{2}\left(2 S_- \bar\rho S_+  - S_+  S_- \bar\rho-\bar\rho S_+ S_- \right) \nonumber \\
&+& \frac{\Gamma_e^{(1)} \bar n_e }{2}\left(2 S_+ \bar\rho S_-  - S_-  S_+ \bar\rho-\bar\rho S_- S_+ \right) \nonumber \\
&+& \frac{\Gamma_n^{(1)} (\bar n_n+1)}{2}\left(2 \sigma_- \bar\rho \sigma_+  - \sigma_+ \sigma_- \bar\rho-\bar\rho \sigma_+ \sigma_- \right) \nonumber \\
&+& \frac{\Gamma_n^{(1)} \bar n_n}{2}\left(2 \sigma_+ \bar\rho \sigma_-  - \sigma_- \sigma_+ \bar\rho-\bar\rho \sigma_- \sigma_+ \right) 
\end{eqnarray}
where $\Gamma_e^{(1)}$ is the decoherence rate of the electron spin ($T_1$ related),  $\Gamma_n^{(1)}$ is the decoherence rate of the nuclear spin ($T_1$ related) and  $\Gamma_n^{(2)}$ is the dephasing of the nuclear spin ($T_2^*$ related). $S_\pm$ are the usual raising/lower operators of the electron while $\sigma_\pm$ are the raising/lower operators of the nuclear spin. $\bar n_e$ ($\bar n_n$ )is the mean photon number of the electron spin (nuclear spin) baths at a temperature T.  

The electron spin also has a dephasing effect, but this cannot be simply modeled by a master equation term of the form $S_z \bar\rho S_z  - \bar\rho$ as the spin bath comes from low frequency noise. Low frequency noise can be modeled in a slightly difference way by considering adding an extra Hamiltonian of the field $H= \lambda f(t) S_z$ to  $\bar H_\text{eff}$ where $ \lambda$ is the coupling strength and $f(t)$ is a symmetric classical normalized Gaussian noise function. The master equation can be solved using standard techniques and the fidelities of the operations calculated.


\begin{thebibliography}{33}
\bibitem{footnote1} While magnetic field manipulation is possible, it is difficult to achieve with extremely high fidelity.
\bibitem{exchange} As we are operating with a magnetic field at $B=50\,\text{mT}$ we are far detuned from these resonances and so will be operating again in the dispersive regime.
\bibitem{footnoten} The technique used to perform the nuclear spin rotations can also be used perform controlled nuclear spin rotations. For instance, if we prepare the electron spin in the 
$\frac{1}{\sqrt{2}} \left[|0\rangle_e+| +1\rangle_e \right]$ with the nuclear spin in the \ketn{0} state. Applying a drive field of frequency $\nu = \Apara-B\ngyro+\frac{\Aperp^2}{2D+2B\egyro}$ for a time t such that 
$\frac{\Omega_0}{4}\left|\frac{\Aperp}{D+B\egyro}-\frac\ngyro\egyro\right| t =\pi/2$ will result in an entangled state of the form $\frac{1}{\sqrt{2}} \left[|0\rangle_e |0\rangle_n+ e^{i A_\parallel t} | +1\rangle_e | 1\rangle_n \right]$. The phase factor can in principle be easy removed, however we do have an issue associated with $T_2^*$ of the electron spin and the time to perform the gate. The time to perform the gate is on the order of $7\mu {\rm s}$ while $T_2^* \sim 90 \mu {\rm s}$. Without composite pulses sequences to enhance this coherence time, we would have an extra intrinsic loss in fidelity of approximately $0.003$.
\end{thebibliography}
\end{document}